# Foaming properties of protein/pectin electrostatic complexes and foam structure at nanoscale.

*Schmidt I[1], Novales B.[1], Boué F.[2*] and Axelos M.A.V[1].*

[1] UR1268 Biopolymères Interactions Assemblages, INRA, F-44300 Nantes, France

[2] Laboratoire Léon Brillouin, UMR 12 CNRS/CEA-IRAMIS, 91191 Gif-sur-Yvette Cedex, France.

* Corresponding author, email address: francois.boue@cea.fr

**Abstract**

The foaming properties, foaming capacity and foam stability, of soluble complexes of pectin and a globular protein, napin, have been investigated with a "Foamscan" apparatus. Complementary, we also used SANS with a recent method consisting in an analogy between the SANS by foams and the neutron reflectivity of films to measure *in situ* film thickness of foams.

The effect of ionic strength, of protein concentration and of charge density of the pectin has been analysed. Whereas the foam stability is improved for samples containing soluble complexes, no effect has been noticed on the foam film thickness, which is almost around 315 Å whatever the samples. These results let us specify the role of each specie in the mixture: free proteins contribute to the foaming capacity, provided the initial free protein content in the bulk is sufficient to allow the foam formation, and soluble complexes slow down the drainage by their presence in the Plateau borders, which finally results in the stabilisation of foams.

**Introduction**

Food macromolecules, such as proteins and polysaccharides, play an important role in the stabilization of foams [1, 2]. They act by retarding liquid film drainage and by producing a viscoelastic layer at the bubble surface which protects the film against rupture and prevents or retards Ostwald ripening. Proteins give a good foamability and high foam stability through their hydrophobicity and possible conformational rearrangements which allow rapid adsorption at the air-water interface leading to the formation of a coherent elastic adsorbed layer [3, 4]. Many polysaccharides being hydrophilic, they do not adsorb at the interface. It has been shown however that they can enhance the stability of foam proteins by a thickening or a gelling effect of the aqueous solution [5, 6]. Some studies have also evidenced an additional role of polysaccharides at the interfacial film [7].

Polysaccharides can interact with adsorbed proteins to form protein-polysaccharide complexes which can increase both the rigidity of the interface and the surface activity of the protein [8]. Aqueous mixtures of proteins and polysaccharides can exhibit various phenomena including complex coacervation, miscibility and segregation [9, 10]. Complex coacervation mainly occurs below the protein isoelectric point as a result of net electrostatic interactions between the biopolymers carrying opposite charges and implies the separation of two phases, one rich in complexed biopolymers and the other phase depleted in both [11]. Recently, the structure of complexes in a region of the phase diagram of an aqueous mixture of pectin and a protein, lysozyme, was described in detail [12]; globular complexes were characterized and showed a strong similarity with aqueous mixtures of lysozyme with a synthetic polymer, sodium polystyrene sulfonate [13].

The combined presence of proteins and polysaccharides may be interesting in foams, particularly in the case of a net attraction between the two kinds of macromolecules. Previous studies on the effect of polysaccharides on foaming properties of food proteins have shown that foam stability is strongly increased [14].

Such protein-polysaccharide interactions would tend to retard film drainage, prevent against Ostwald ripening and strongly increase the interface stability [15].

The present study deals with the foaming properties of napin/pectin mixtures. Pectin is a natural polysaccharide extracted from citrus and apple fruits at the industrial scale. It is widely used in the food industry for its gelling and stabilising properties. This polymer is constituted by a galacturonic acid backbone with 1-4 linkages interrupted by L-rhamnose residues; the side-chains are constituted by neutral sugars, mainly D-galactose and L-arabinose [16]. The number of galacturonic acid monomers esterified with methanol for 100 monomers is defined as the degree of methylation (DM): DM = [COOH]/([COOCH$_3$+COOH]).

The degree of charge is determined by (100 – DM)/100.

Under certain well-defined conditions, pectin can form viscoelastic solutions and structured networks, a property that is widely exploited in jams and marmalades [17].

The 2S rapeseed protein, called napin, was extracted by successive chromatographic stages [18]. This protein is characterised by its strong basicity (pI = 10-11) and its low molecular weight (14 000 Da) [19].

Napin being a positively charged globular protein, and pectin being a negatively charged semi-flexible polysaccharide, they form electrostatic complexes. These complexes have been characterized in part in the bulk of aqueous mixtures [20], and show some similarities with pectin-lysozyme complexes of reference [12]. In foams, the role of these complexes on the foaming properties has to be precised. In this paper, we will analyze the effect of DM (or charge densities of the polysaccharide) and of ionic strength on the foaming properties at pH 7 of such complexes. The foaming properties will be analyzed with regard to foaming capacity, foam stability but also with regard to *in situ* foam film thickness. The *in situ* film thickness of foams is determined by using a method based on an analogy between SANS by foams and the neutron reflectivity of films. This method has been developed by Axelos and Boué [21] and recently applied by Ropers and al. [22] to polysaccharide/surfactant systems for the first time. Based on these different characteristics of foams, the role of each specie present in the mixture (free proteins or protein/pectin complexes) will finally been precised.

**Material and Methods**

**Material**

The napin used in this paper was extracted from rape meal, isolated and purified according to the method developed by Berot et al. [18]. Its physicochemical characteristics were determined previously [19]. Its hydrodynamic radius determined by dynamic light scattering has been measured at 1.98 nm and its charge is 10.5 at pH 7.

Two pectins with different degrees of methylation, 43% and 74%, were used which will be noted respectively, DM43 and DM74. Pectins have been graciously provided by Copenhagen Pectin. The samples were purified by precipitation in ethanol acidified by hydrochloric acid as indicated in previous papers in order to remove traces of divalent cations [23].

**Methods**

*Preparation of samples*

Prior to mixture, solutions of pure pectin at 2g/L and of pure napin at 10, 2, or 0.2g/L were prepared by dissolution in a same phosphate buffer to maintain the pH at 7. All solutions were prepared by using MilliQ water (18,2 MΩ.cm.). The buffer, a mixture of an aqueous solution of $KH_2PO_4$ and of an aqueous solution of $Na_2HPO_4, 2H_2O$, was set to have an ionic strength of 25mM or 148mM. For SANS experiments, the buffers have been prepared by dissolution of salts in a pure solution of $D_2O$ (Eurisotop, France) in order to obtain a maximal contrast.

Solutions were homogenized with a vortex for a few seconds; then they were gently stirred for 2h at room temperature and at least for 12h at 4°C to complete solubilization. Two equal volumes of each of the solutions were then mixed, which led to a concentration of 1g/L of pectin and 5, 1 or 0,1g/L of napin in the final mixture. Solutions were homogenized with a vortex for a few seconds and then stored at 4°C and used in a delay of 7 days.

The concentration of napin in the sample was determined by UV spectroscopy using a UV-vis Shimadzu UV-1605 spectrometer in a $\lambda$ range from 250 to 700 nm. The titration was done for the absorption maximum at a wavelength of 280nm. The concentration of pectin in the sample was measured by colorimetric titration [24].

*Solubility diagrams and characterization of complexes size.*

Phases were identified by visual inspection. The samples were observed every day and a stable state was observed after 7 days. The mixtures form either monophasic or diphasic phases. These phases are either clear, turbid, or form a precipitate with a supernatant either clear or turbid [20]. The stability of the sample was checked by measuring optical density at 600 nm of monophasic samples or of the supernatant of biphasic samples. This wavelength was chosen because it is different from the wavelengths where aromatic acids of the protein absorb. We determined that the limit between clear and opalescent samples could be distinguished by eyes for an optical density of 0.05.

Dynamic light scattering was performed using a Nanosizer ZS (Malvern Instruments, UK) in order to determine the hydrodynamic radius of the complexes. The instrument was used in the backscattering configuration where detection is done at a scattering angle of 173°. Dilute solutions were measured in a 1 cm path-length spectroscopic plastic cell at 20°C. The hydrodynamic radius was measured in triplicate. Each measurement corresponded to three autocorrelation functions recorded during 90s. Before using the Malvern Nanosizer, its range of reliability has been checked by comparing the results with data obtained by static light scattering.

*Foaming properties*

Foaming experiments were conducted on a "Foamscan" apparatus developed by IT Concept (Longessaigne, FRANCE). With this instrument, the foam formation, the stability and the drainage of liquid from the foam were followed by conductimetric and optical measurements. After calibration of the conductimetric electrodes, 8mL of solution in the sample cell were sparged with gaseous $N_2$ through a porous glass filter at a flow rate of 25 mL/min. All foams were allowed to reach a final volume before draining of 35 mL after which gasflow was stopped and the evolution of the foam was analysed. The foamability corresponds to the time needed to reach 35 mL of foam volume. After stopping the bubbling, the free drainage of the foam was followed for 1200s. Several parameters were automatically recorded by the "Foamscan" analysing software. The drainage of liquid from the foam is followed via conductivity measurements at different heights of the foam column. A pair of electrodes at the bottom of the column was used for measuring the quantity of liquid that was not in the foam, while the volume of liquid in the foam was measured by conductimetry with three pairs of electrodes located along the glass column. The volume of foam is determined by use of a CCD camera (Sony Exwave HAD).

*Interfacial properties*

Interfacial properties of napin and of napin/pectin mixtures in solution were studied by measuring the surface tension as a function of time. The measurement was performed with an automated drop tensiometer (IT Concept, Longessaigne, France) in the rising drop configuration. Temperature was kept constant at 20°C. An air bubble (7.6µL) was formed in a cuvette containing the sample solution. The bubble was illuminated by a uniform light source and its profile was imaged and analysed by a CCD camera. Evolution of the surface tension was determined with time by bubble shape analysis using Laplace's equation. The decrease in surface tension was followed and dynamic oscillations of the bubble (area change of 5.5%) were applied to determine the dilatational modulus [25].

*Small Angle Neutron Scattering (SANS) on foam samples*

SANS measurements were performed on PAXY spectrometer (Orphée reactor, LLB, CEA Saclay, France) according to the method developed by Axelos and Boué [21]. They have shown that specific characteristics of foams can be determined:

- the mean size of bubbles can be determined at low q,

- the thickness of films at intermediate q

- the structures inside the Plateau borders at large q.

This method has been explained more in detail by Ropers and al. [22].

In the present study, two different configurations (wavelength/sample detector distance) were used to cover a range of scattering vectors q lying between $4.34 \ 10^{-4}$ and $5.22 \ 10^{-2}$ Å$^{-1}$ in configuration 1 and between $1.22 \ 10^{-3}$ and $1.08 \ 10^{-1}$ Å$^{-1}$ in configuration 2.

Foams prepared from samples containing 1g/L of protein and 1 g/L of pectin were analysed. The foams were prepared in a foam cell, or column, very similar to that described by Axelos and Boué [21]. The lower part of the cell is constituted of a quartz tube, which can be diametrally crossed by the neutron beam on a wide range of height. It is similar to the glass cylinder of the Foamscan apparatus, with a larger diameter in order to increase the intensity arising from a higher number of foam films met by the beam (30 mm wide against 19 mm in Foamscan). The bubbling was adjusted to get a foam height of 10 cm. As soon as it was stopped, the neutron scattering intensity was recorded on the free-draining foam every 60 s for napin foams and 120 s for napin/pectin foams.

*Small Angle Neutron Scattering (SANS) data analysis*

First SANS standard corrections for sample volume, neutron beam transmission, and incoherent scattering due to protons or deuterons were applied. The scattering of the empty cylinder was subtracted from the scattering of the cylinder filled with the foam and this difference was divided by the scattering intensity of 1 mm of water. Data configuration were corrected from the solvent contribution by subtracting a constant which takes into account the $D_2O$ contribution and the incoherent scattering due to the protons of napin and Pectin.

Second, a software, available at the LLB for analysis of reflectivity measurements, was used here to fit the SANS data. The analogy of SANS from a foam with multiwall reflectivity, and usual reflectivity by a thin layer, has been formerly evoked [21, 26, 27]. It is assumed that a foam submitted to an incident radiation beam is akin to a collection of M mirrors (the M foam cell walls). The neutron parallel beam (with an angular collimation $\delta\theta_{coll}$), hits each mirror m with an incidence angle $\theta_m/2$, and undergoes a "specular" reflection such that the total deflection is a scattering angle $\theta = \theta_m$. This is sketched in **Figure 1**.

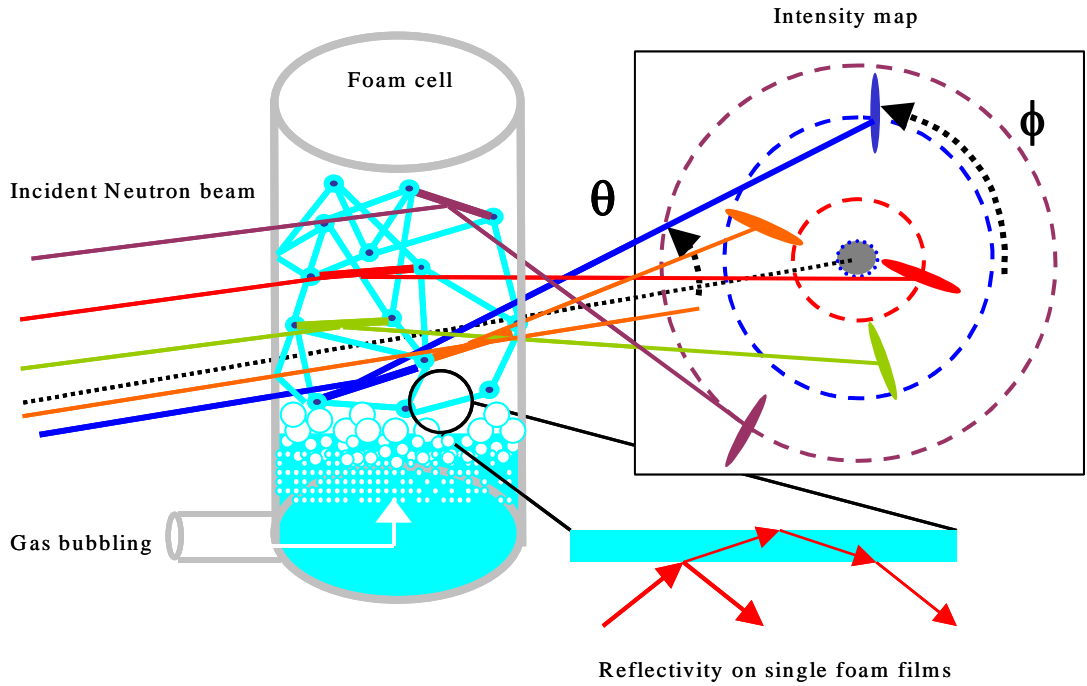

**Figure 1**. Sketch of foam produced by bubbling in a cell, with incident and reflected beams on some of the cell walls. θ is the angle between incident and reflected beam, and ϕ is the azimuthal angle (both angles shown here for the blue ray).

Several walls may deflect the beam by the same angle θ, at different angles φ; then the reflection spots are located on the same circle centered on the trace of the direct beam on the detector. Radial averaging along this circle gives the scattered intensity for the corresponding scattering vector of modulus $q = (4\pi/\lambda)\sin(\theta_m/2)$. The result is equivalent to what obtained with a single mirror oriented at several angles $\theta_m$. In practice, the observed distribution of the intensity is continuous, not spot-like. This is due to the fact that each reflection spot is actually spread into a "spike" of large angular width $\Delta\theta_m$. This has been attributed to the fact that foam cell walls are slightly bent [21] as seen also on single films [26]. If the effective incident angle distribution is even wide and flat enough (this has to be kept in mind), the result is close to a reflectivity curve obtained by rotating continuously a single mirror in front of the beam.

When such reflecting cell walls are a thin layer of water (10 nm < t < 50 nm), we can use a simple simulation software (available for example at www-llb.cea.fr). If it was a single dioptre, i.e. an infinite layer of species A with interface with vacuum, the rate of reflected intensity R(q) shows, as a function of θ or better of q, two zones: 1) For $q < q_C$, a "reflectivity plateau". Reflectivity is total (R(q) = 1) because there is no refracted beam (only an evanescent wave). The upper edge of the plateau is the critical value; $q_c \sim (4\pi Nb)^{1/2}$ where Nb is the scattering length density of the medium. 2) For $q > q_C$, there is a refracted beam, so only a part of the beam is reflected, and R(q) decays fast, with asymptotic variation $q^{-4}$ (Fresnel law).

When the layer thickness t becomes finite, fringes appear in the decaying part, with maxima separated by a period $\Delta q = 2n\pi/t$. This is related to interferences existing between the fractions of the beam reflected at each film-vacuum interface. For larger thicknesses values, this effect coexists with Descartes-Fresnel law, and the reflectivity plateau is not modified, i.e. $q_C$ is the same. When the thickness decreases, the evanescent wave reaches the other interface, so a part of the beam is refracted. The first maximum tends asymptotically to $q = 2\pi/t$ when the thickness t tends to zero.

These effects, which are smoothed by the angular width of the beam, can be magnified graphically using a $q^4 R(q)$ representation. In this representation, when a full reflectivity plateau exists for R(q) at low q, $q^4 R(q)$ increases like $q^4$ in this region. When q becomes larger than $q_C$, R(q) starts decreasing, which results in a first maximum in $q^4 R(q)$, observed at $q_C$. This maximum does not correspond to an interference fringe. At larger q, $q^4 R(q)$ reaches a second maximum which corresponds to the first interference fringe; the following maxima at larger q have the same origin, i.e. fringes, and should be separated by $\Delta q = 2\pi/t$.

Reference [22] discusses and shows a Figure (Figure 4 in [22]) of a series of reflectivity curves at decreasing thickness, for the same Nb ($D_2O$). Still in Reference 22 (Figure 5 in [22]) is shown the position of the first maximum versus the thickness of the film between 50 and 500 Å. It can be concluded that at large thicknesses (over 300Å), the position of the first maximum of $q^4R(q)$ does not depend on the thickness and is very close to $q_C$. At low thicknesses (t < 300 Å, see for example 125 Å), this "$q_C$ maximum" disappears.

**Results**

*Solubility diagrams and characterization of complexes*

The solubility diagrams for napin/pectin mixtures are given **Figure 2**. Mixtures of napin and pectin formed either monophasic or diphasic states. As the concentration of napin increases, these states were successively clear, turbid, or formed a precipitate with a supernatant either clear or turbid. These observations show that napin and pectin interact together.

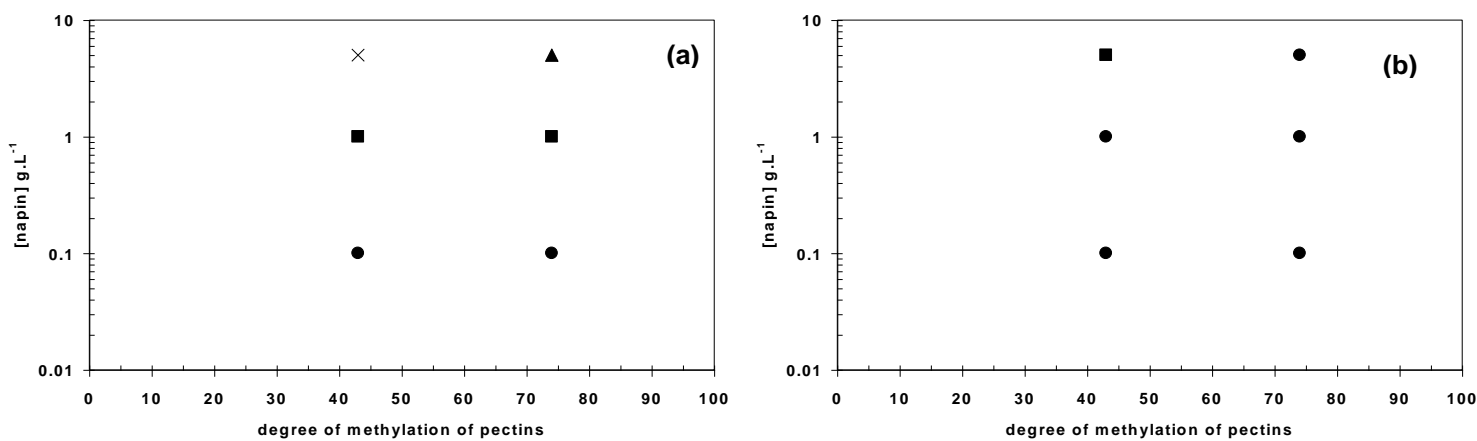

**Figure 2**: Partial solubility diagrams of mixtures of pectins DM43 and DM74 (1g.L$^{-1}$) with napin (0.1, 1 and 5g.L$^{-1}$) at I= 25mM (a) and I=148mM (b)(●) clear; (■) turbid; (✕) precipitate with clear supernatant; (▲) precipitate with turbid / opalescent supernatant.

At an ionic strength of 25mM, the samples were successively clear ([napin] = 0,1g/L), turbid ([napin] = 1g/L) and forming a precipitate with a supernatant either clear or opalescent ([napin] = 5g/L). The supernatant was clear for the mixture containing DM43 and opalescent with DM74. At an ionic strength of 148mM, all samples were clear, except for the mixture DM43/Napin with a concentration of 5g/L of napin which was turbid.

Therefore, the phase behavior differs according to the ionic strength: napin and pectin interact for lower concentrations of napin at 25mM of ionic strength compared to mixtures at 148mM of ionic strength. Results for concentration of free napin in the solutions determined by UV spectroscopy showed that at 148 mM and 1 g/L of protein, 78% of napin were free whatever the DM while at 25 mM, only 4% of napin were free for pectin DM43 and 9% for pectin DM74. Light scattering results indicated that at 25 mM, the aggregates were much bigger than at 148 mM. For limpid solutions, the aggregate size was between 200 and 500 nm. For turbid solutions, an aggregate size larger than 1500 nm was observed.

*Foaming properties*

In order to study the foaming and interfacial properties of napin-pectin mixtures, conditions to obtain monophasic samples were chosen. The concentration of pectin was set at 1g/L. Protein concentration was fixed at 0.1, 1 and 5 g/L. The foaming properties were studied by measuring the time evolution of the volume of foam and the volume of liquid in the foam for pure solutions of napin or for napin/pectin mixtures.

We checked that solutions of pure pectin do not foam for concentrations as high as 10g/L which is ten times higher than the concentration set in this paper. Thereby, all foaming properties will arise from napin or napin/pectin complexes.

We will determine the role of each specie on the final foaming properties and analyse the foaming properties with regard to foam formation and to foam stability.

*Foam formation*

At 0.1 g/L of protein, we observed a difference of aspect during the foam formation according to the ionic strength. At 25 mM, the foam exhibited big bubbles which rapidly dried and collapsed. At 148 mM, wet foam with relatively small bubbles was obtained.

**Figure 3** shows an example of foam obtained at a) 25 mM and b) 148 mM. These foams were produced by bubbling a solution on a thin glass column of width 2 mm. This made it possible to acquire images of foam bubbles with a sufficient contrast to observe their size. These images clearly show the difference between foams obtained at these 2 ionic strengths. The foam obtained at 25 mM is composed of larger bubbles (0.8 to 3 mm) than at 148 mM (0.3 to 1.5 mm).

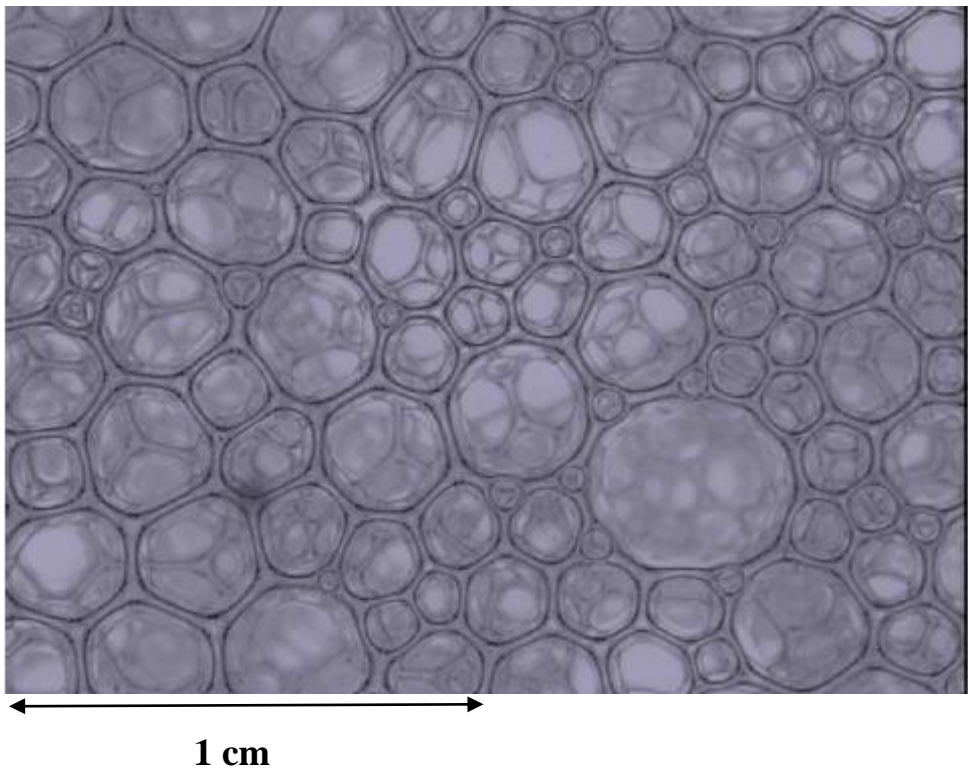

**1 cm**

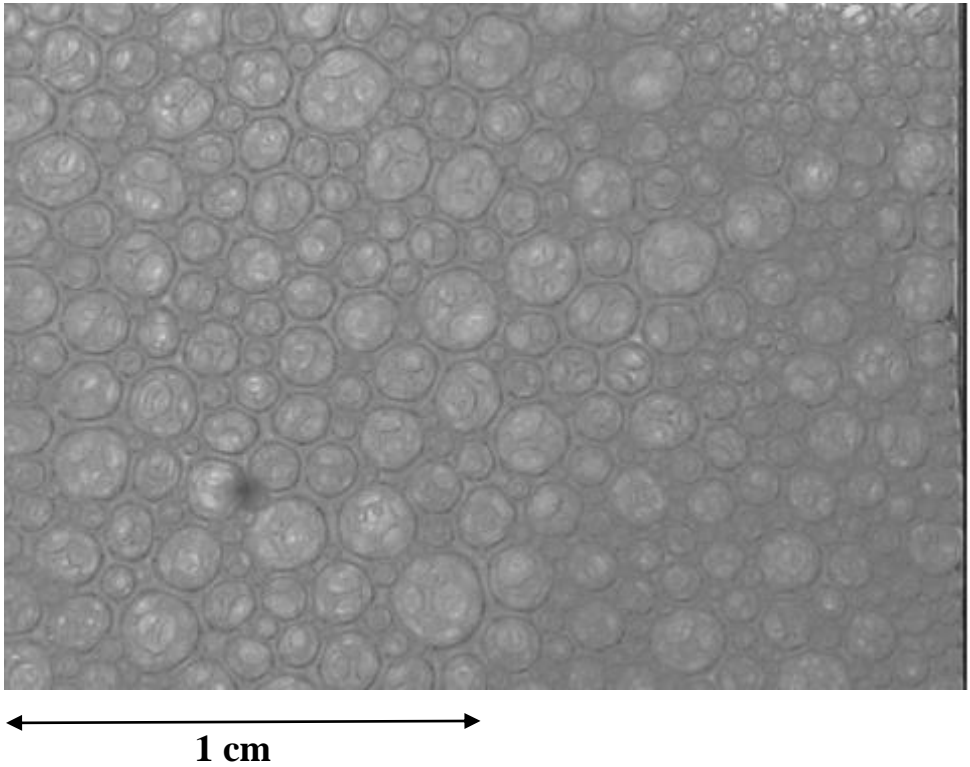

**1 cm**

**Figure 3:** Photos of a foam area of 2 cm x 1.5 cm. Pectin concentration 1g/L, protein concentration 0.1 g/L. **a** (Top): Ionic strength 25 mM. **b** (Bottom): ionic strength 148 mM.

At 1 g/L and 5 g/L of protein, no difference of aspect was noticed between foams prepared at 25 or 148_mM of ionic strength: the foams were white, with small spherical bubbles having a diameter of nearly 0.3mm. The addition of pectin did not affect the foam formation.

Two parameters were distinguished:

- the maximal foam volume that was initially set at 35mL; when this volume was reached, the bubbling was stopped and the free drainage of the foam was followed during 1200s

- the effective maximal foam volume that was observed.

The time to reach the value of the maximal foam volume before draining (set at 35 mL) was the same, 70 s, for pure napin solutions and for mixtures, whatever the ionic strength. This indicates that all samples have the same foam capacity.

The effective maximal foam volume reached is 38mL, as seen on **Figure 4.a**. The time to reach this value is the same, 105 s, for pure napin solutions and for mixtures whatever the ionic strength. This confirms that all samples have the same foam capacity. The fact that the effective maximal foam volume (38mL) is higher compared to the maximal foam volume initially set at 35mL reveals very good foaming properties.

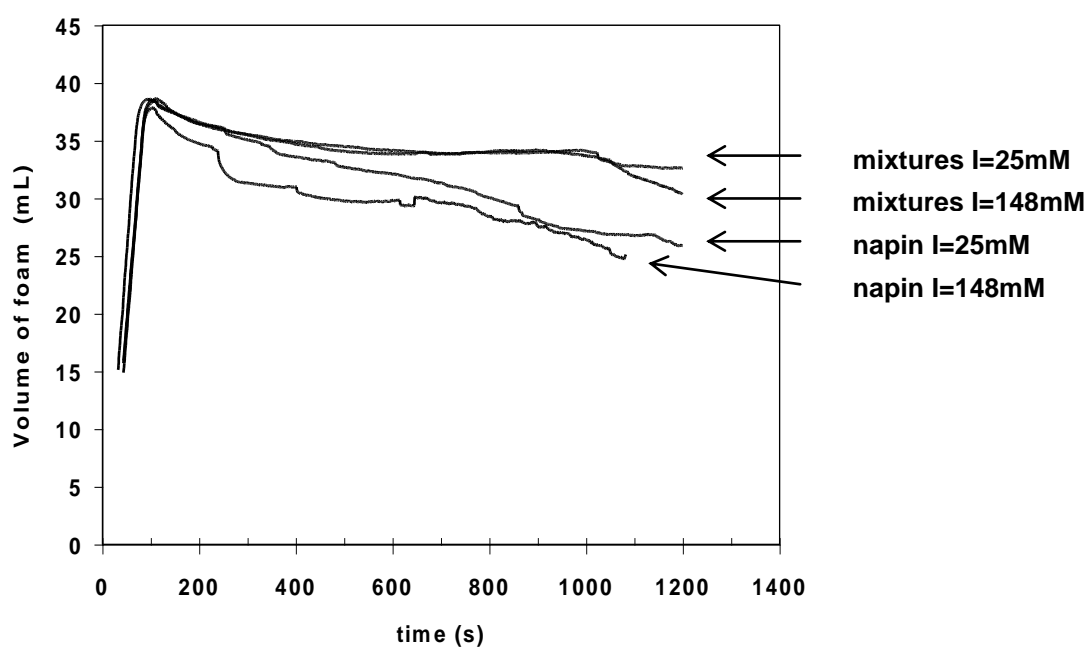

**Figure 4.a**: Evolution of the volume of foam from napin ($1g.L^{-1}$) and from mixtures containing complexes of napin ($1g.L^{-1}$) complexed with pectin DM74 ($1g.L^{-1}$) at 25mM and 148mM of ionic strength.

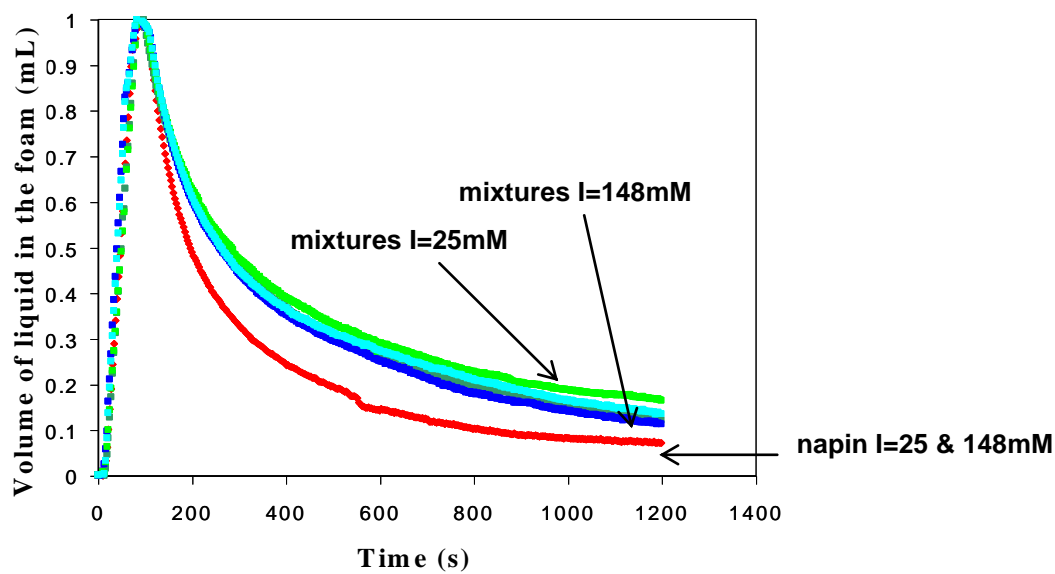

**Figure 4.b** Evolution of the volume of liquid in the foam from napin (1g.L$^{-1}$) and from mixtures containing complexes of napin (1g.L$^{-1}$) complexed with pectin DM74 (1g.L$^{-1}$) at ionic strengths 25mM and 148mM. For all curves, the maximum value of the volume after the end of bubbling (which varied between 5.5 and 6.5 mL) has been renormalized to one.

*Foam stability*

The major differences between foams of pure napin and foams of napin/pectin mixtures appeared in the foam destabilization. At 5 g/L of napin, the foam volume measured 20 minutes after the end of bubbling was above 30 mL, indicating that these foams were very stable. No differences were noticed between the different samples. At 1g/L and 0.1 g/L of protein, the effect of pectin is on the contrary very clear: it increases the foam stability. This is illustrated in **Figure 4.b** for a napin concentration of 1 g/L: it indicates clearly that foams produced from pure napin solutions have a faster drainage and are thereby less stable compared to foams produced from napin/pectin mixtures. The initial slopes of the drainage curves are higher for pure napin solutions than for mixtures at 1 g/L. Ten minutes after the end of foaming, less then 10% of liquid is remaining in the foam obtained with pure napin solutions. In presence of pectin, the liquid volume in the foam is approximately 35 % at the same time. As a dryer foam is more sensitive to coalescence and collapse, the consequence of this difference in liquid content in the foam was that after 20 min, the foam volume observed for pure napin solutions was lower than for mixtures, whatever the protein concentration. The ratio of free protein/complexes was determined in the foam and in the drained liquid by the method of Markwell [28](a modification of the Lowry method of protein determination by the addition of sodium dodecyl sulfate in the alkali reagent and an increase in the amount of copper tartrate reagent to simplify protein determination). These measurements were only possible at an ionic strength of 148mM. At 25 mM of ionic strength, drainage was too fast and the liquid volume incorporated in the foam was too small to allow the measurement. **Table 1** gives the respective percentages of proteins in the drained liquid and in the foam. We observed that almost half of the protein content was found in the foam at the end of the experiment.

| Sample | | Volume (mL) | Protein concentration (mg/mL) | Mass (mg) | Protein content (%) |
|---|---|---|---|---|---|
| Napin + DM 43 | Drained liquid | 6.12 | 0.83 | 5.080 | 56% |
| | Foam | 2.04 | 1.90 | 3.920 | 44% |
| | Total | 8.16 | 1.10 | 8.976 | 100% |
| Napin + DM 74 | Drained liquid | 5.56 | 0.82 | 4.559 | 51% |
| | Foam | 2.55 | 1.71 | 4.365 | 49% |
| | Total | 8.11 | 1.10 | 8.921 | 100% |

**Table 1:** Composition of foams and of drained liquids regarding the total protein content at 148mM of ionic strength

| Sample | | Free / Complexed proteins | Protein concentration (mg/mL) | Mass (mg) | Protein content (%) |
|---|---|---|---|---|---|
| Napin + DM43 | Drained liquid | Free | 0.65 | 3.966 | 78% |
| | | Complexed | 0.18 | 1.09 | 22% |
| | Foam | Free | 1.49 | 3.04 | 78% |
| | | Complexed | 0.43 | 0.885 | 22% |
| Napin + DM 74 | Drained liquid | Free | 0.66 | 3.686 | 81% |
| | | Complexed | 0.16 | 0.873 | 19% |
| | Foam | Free | 1.28 | 3.276 | 75% |
| | | Complexed | 0.43 | 1.089 | 25% |

**Table 2**: Composition of foams and of drained liquids regarding the protein concentration at 148mM of ionic strength; the table indicates the free and the complexed protein content in each fraction.

**Table 2** gives the respective percentages of free proteins and complexes in the drained liquid and in the foam. We determined that approximately 78% of the protein content in the foam was free proteins. This amount of free protein was the same than that observed in the solutions before foaming. These results show that a large part of complexes is still present in the foam at the end of the experiment.

*Interfacial properties*

As the ability to form foams is linked to the adsorption kinetics of surface active molecules like proteins, the interfacial properties of the napin samples and napin/pectin complexes were studied.

*Interfacial properties of napin*

The decrease of surface tension for napin solutions with increasing protein concentration from 0.1 to 3g/L was followed for both ionic strengths I=25 and 148mM.

The surface tension reaches a plateau in 4h indicating that an equilibrium was reached. The decrease of the surface tension is fast for all samples whatever the protein concentration; the higher the concentration, the faster the decrease of surface tension.

No difference of surface tension at equilibrium was noticed with the ionic strength for a same protein concentration (**Table 3**).

| [napin] g.L$^{-1}$ | $\gamma$ equilibrium (mN/m) I=25mM | $\gamma$ equilibrium (mN/m) I=148mM |
|---|---|---|
| 0,1 | 54,1 | 54,2 |
| 0,5 | 52,3 | 52,4 |
| 1 | 49,7 | 50,7 |
| 1,5 | 46,4 | 45,6 |
| 3 | * | 42,2 |

**Table 3**: Surface tension at equilibrium for napin solutions with increasing protein content

The protein concentration of all samples is beyond the critical association concentration (c.a.c. = $5.6 \times 10^{-5}$ g/L), established by Krause and Schwenke [29]. Beyond the c.a.c., the surface tension should be stable. The decrease of surface tension that is still observed for the samples may be attributed to reorganisations of proteins at the interface or to the presence of **multicouches**.

*Interfacial properties of mixtures*

The decrease of surface tension for napin/pectin mixtures at 1g/L of protein and 1g/L of pectin was followed for both ionic strengths. These four samples have the same profile of adsorption kinetic, only one of them is represented on **Figure 5** at 148mM of ionic strength in comparison with the isotherm of napin.

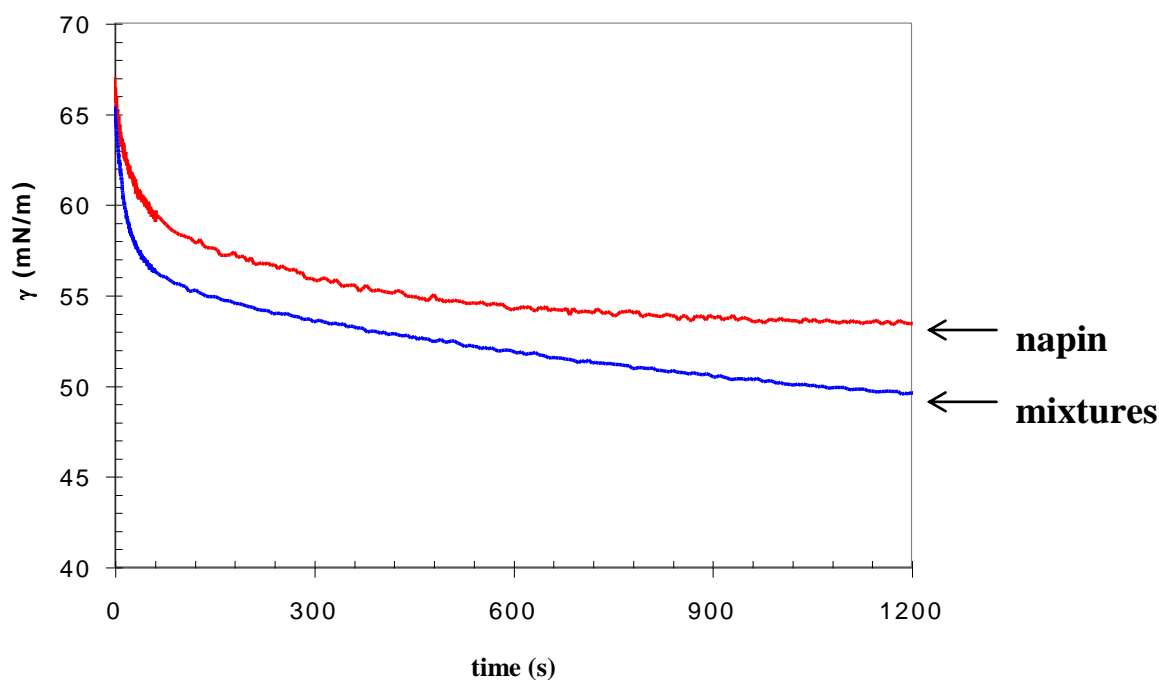

**Figure 5**: Adsorption kinetics for solutions of napin (1g/L, above) and solutions of complexes (1g/L of protein, 1g/L of pectin, below)

As for napin, the decrease of the surface tension for the mixtures is fast and reached a plateau for 15000s. The surface tension at equilibrium (**Table 4**) is lower for mixtures compared to napin solutions, which reveals improved interfacial properties for mixtures compared to napin.

| Sample | γ equilibrium (mN/m) I=25mM | γ equilibrium (mN/m) I=148mM |
|---|---|---|
| P43 1g.L$^{-1}$ + napin 1g.L$^{-1}$ | 46,5 | 45,5 |
| P74 1g.L$^{-1}$ + napin 1g.L$^{-1}$ | 47,3 | 45,5 |
| Napin 1g.L$^{-1}$ | 49,7 | 50,7 |

**Table 4**: Surface tension at equilibrium for napin and for napin/pectin mixtures

Dynamic oscillations of the bubble were applied to determine the dilatational modulus [25]. We focused on the early stages (30s after the bubbling started). The surface tension for complexes is 59mN/m versus 62mN/m for napin, confirming that the interfacial activity of complexes is immediately higher compared to napin. The dilatational modulus (determined for γ = 59mN/m) is 17mN/m for napin versus 24 to 27mN/m for the complexes indicating that the interfaces formed with complexes are more rigid.

*Small Angle Neutron Scattering (SANS)*

The log-log plot of the scattering intensity as a function of q was similar whatever the samples, pure napin solution or napin/pectin mixtures. The curves could be superimposed by shifting the different spectra according to the y axis. **Figure 6** represents the curve obtained for foams stabilized with napin at 1g/L and 148mM of ionic strength.

All spectra were composed of four different parts, as seen in **Figure 6**.

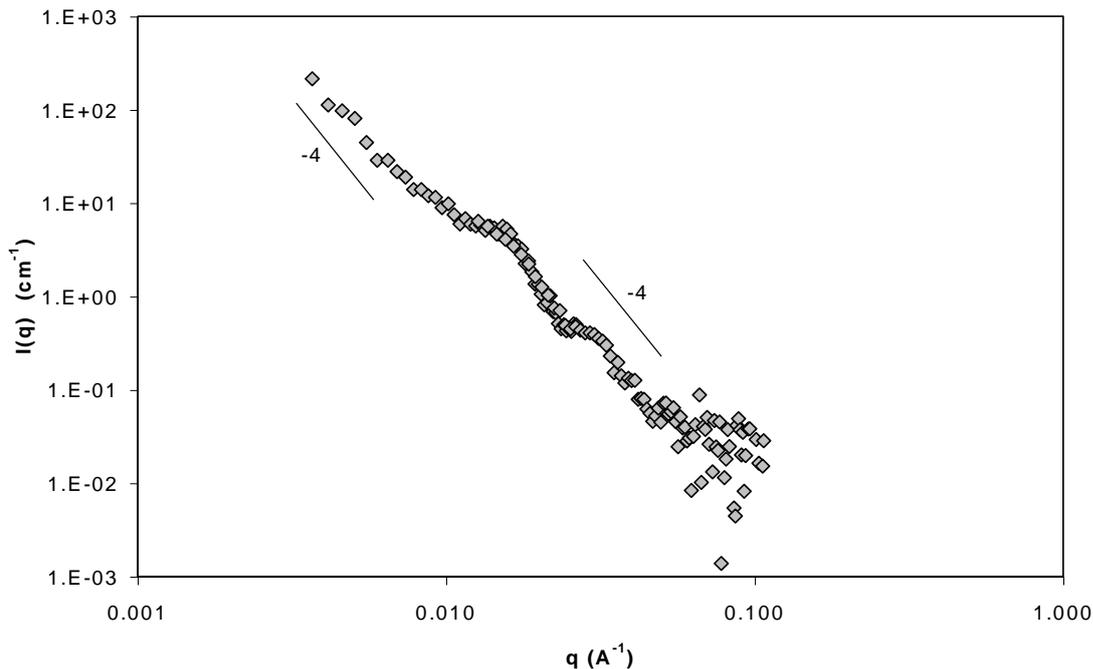

**Figure 6**: log-log plot of the scattering intensity for foams stabilized with napin at 1g/L and 148mM of ionic strength.

At low q, a $q^{-4}$ decay is clearly visible on wet and dry foams. At intermediate q a shoulder is present which appeared during foam drainage. Another bump appeared at large q which is visible on dry and wet foams. Finally, a $q^{-4}$ decay is observed at very large q, with large statistical fluctuations due to a very low scattering intensity.

As all the scattering spectra exhibited a $q^{-4}$ decay, it was interesting to use a $q^4 I(q)$ vs q representation, as shown in **Figure 7**.

On the corresponding curves, two peaks were always visible, which are characteristic of dry foams. The position of these peaks was the same whatever the samples (in absence or presence of pectin); they correspond to the shoulders in the log I – log q plots. The first maximum, at the lowest q (arrow on the left hand), in $q^4 I(q)$ (corresponding to the first shoulder in log-log plot) at intermediate q is observed for $q = 1,75 \times 10^{-2}$ Å$^{-1}$ whereas the second maximum, at higher q (arrow on the right hand side), is seen for $q = 3,31 \times 10^{-2}$ Å$^{-1}$.

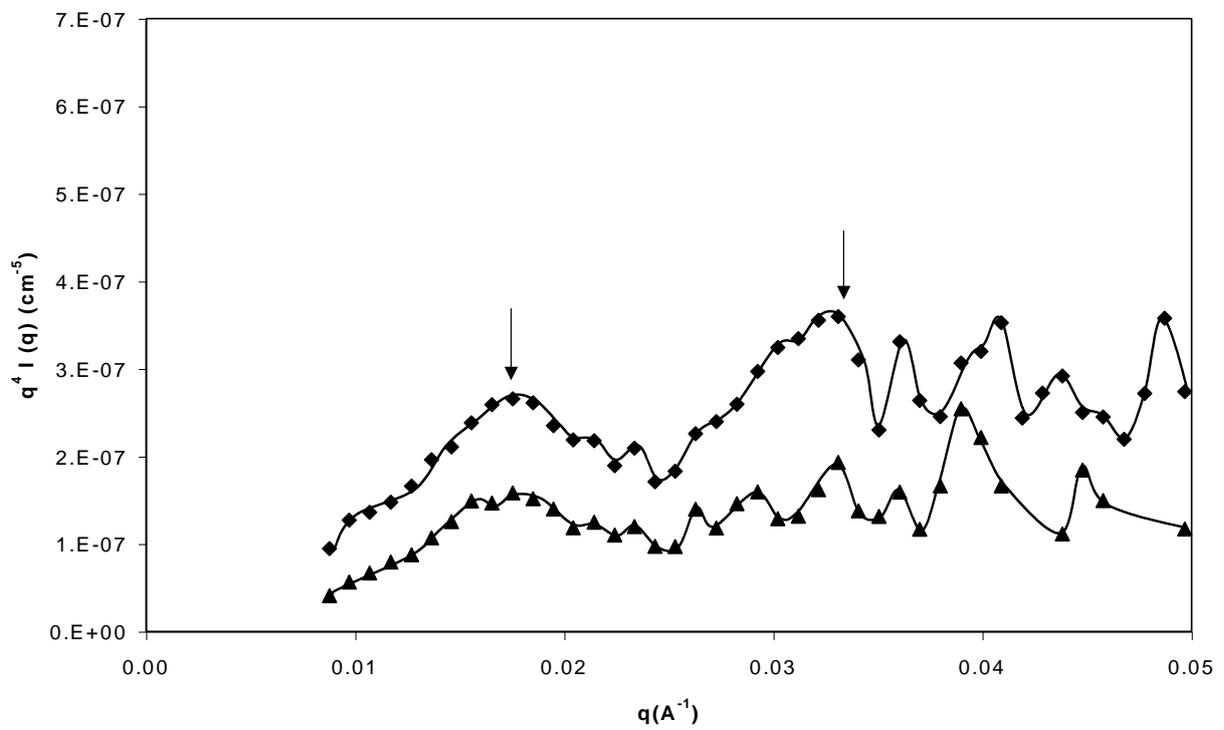

**Figure 7**: Plot of $q^4I$ vs q for dry foams stabilized with napin at 1g/L (♦, diamonds, above) and with mixtures of napin at 1g/L with pectin P74 at 1g/L (▲, triangles, below) at 148mM of ionic strength.

In classic analysis of small-angle neutron scattering, a $q^{-4}$ behavior is characteristic of the scattering by interfaces with well-defined surfaces in a biphasic system [30], with

$$I\,(\text{cm}^{-1}) = 2 \cdot \Delta\rho^2\,(S/V)\,(1/q^4)$$

where $\Delta\rho^2$ is the contrast, *S* the total surface, *V* the total volume, and *S/V* the specific area. The system is diphasic here, since the foam can be considered to be a porous medium of solution with air pores [21]. More precisely, the foam is a more or less dense packing of bubbles with two types of surfaces: the surface between the air and the Plateau borders and the two parallel surfaces which face each other in the films. These surfaces are very sharp (roughness less than a few Angströms), giving rise to the Porod law behavior.

In these conditions, these peaks are characteristic of the mean thickness of films in the foams. The thickness of the film was determined with a fit usually used for reflectivity experiments data and which consists mainly in a fit of the peak position. The thickness of films could not be determined at 25mM of ionic strength because the foams were not stable enough. At 148mM, the thickness of films in foams produced from pure napin solutions was 320 to 350Å and around 315 Å in foams produced from napin/pectin mixtures, whatever the DM of pectins. SANS experiments indicated thereby that the thickness of films in the foams is more or less the same whatever the sample, pure napin or napin complexed to pectins.

**Discussion**

The comparison between foams of pure napin solutions and foams of mixtures clearly indicates that adding pectins to solutions of napin improves the foaming properties. The results also indicated that pectin-napin interactions lead to the formation of complexes which are still present in the foam in a large quantity at the end of the experiment.

*Effect of protein concentration*

As the concentration of pectin has been set at 1g/L, varying the concentration of napin (or positive charges) leads to vary the +/- charge ratio, which has an impact on napin/pectin interactions, as can be seen on the solubility phase diagrams on figure 1. Increasing the protein concentration in the mixture favours napin/pectin interactions which can lead to precipitation for 5g/L of napin and an ionic strength of 25mM. The observation or not of precipitation can be related to the size of complexes determined by light scattering: for 25mM of ionic strength, the size of complexes is around 1500 nm, which results in precipitation, whereas at 148mM of ionic strength, their size is 200-500 nm, which results in soluble complexes. Electroneutrality boundary is also an important parameter for the foaming properties. In fact, it has been reported on surfactant/polymer systems that the best foaming properties are obtained at the boundary between soluble and insoluble complexes [31].

At 5 g/L of protein, for both ionic strengths, the foam obtained for pure napin solutions was so stable that it was difficult to see an improvement of the foaming properties in comparison with napin/pectin mixtures. The behaviour of napin is predominent in these conditions.

However, we saw a slower drainage for mixtures which indicates that complexes contribute to the foam stabilization.

At lower protein concentration, on the contrary, we observed that complexes were efficient to improve the foaming properties of napin:

- at 1g/L and 0.1g/L of protein, complexes did not affect the foam formation but they improved the foam stability. The free and complexed protein content was determined before and after foaming. It indicated that at 25mM of ionic strength most of the proteins were complexed before foaming (91 to 96%) but drainage was too fast and the liquid volume incorporated in the foam was too small to allow measurement after foaming.

At 148mM of ionic strength, only 22% of the proteins were complexed before foaming but drainage was slowed down. After foaming, the free and the complexed protein contents were determined respectively around 78% and 22%. These results evidence an impact of the ionic strength on the foaming properties and on the protein/pectin interactions.

*Effect of ionic strength*

The influence of ionic strength on protein-pectin interactions is evidenced by the solubility diagrams which indicate that precipitation occurs at low ionic strength, near electroneutrality, whereas the solution stays monophasic at high ionic strength. This implies that electrostatic interactions are predominant and that their strength can be modulated by the buffer concentration. It is illustrated by the results of protein concentration determination: at low ionic strength, only 4% of napin is free which is due to the low screening effect of buffer ions whereas at high ionic strength, 78% of napin is free, linked to a higher screening effect by buffer ions.

The range of electrostatic interactions concerns not only the interactions between pectins and napins but also between complexes already formed which can interact and lead to much bigger structures, as illustrated by the 1500nm size of complexes determined by light scattering at 25mM. This effect is more limited at high ionic strength, which leads to smaller structures (size of 200-500nm). We cannot definitely exclude at this stage an effect of chelation of the phosphate ions by the cationic ligand sites of napin. Such chelation could identically explain why the extent of pectin/napin aggregation was reduced as the buffer concentration increased. An accurate discussion is beyond the scope of the paper.

At 1g/L, with regard to the formation of foams, no influence of ionic strength has been noticed: all samples reach the maximum foam volume (initially set at 35mL) in 70s.

On the contrary, with regard to the stability of foams, influence of ionic strength is noticed. stability could actually be determined at high ionic strength only, because of a very fast drainage and insufficient liquid volume in the foam at low ionic strength. This fast drainage at 25mM can be related to the bigger size of complexes which favour precipitation and loss of stabilisation of foam interfaces by complexes. In this case (low ionic strength), the free napin content (4%) is not sufficient to stabilize the interfaces.

At 0.1g/L of protein, ionic strength creates a difference in the foam formation. At 25mM of ionic strength, the foam could not form because it collapsed rapidly, which must be due to the small amount of free protein in the sample, not sufficient to stabilise the foam during its formation. At 148mM of ionic strength, the foam formation was possible thanks to a sufficient amount of free protein in the sample. These results let us conclude that a minimum concentration of free protein in the sample is required to allow the foam stabilisation during its formation.

*Effect of polymer charge density (DM of pectins)*

The effect of the charge density of the pectins has only been observed for the solubility diagrams determination. It has been noticed that phase transition between clear and turbid or monophasic and biphasic samples occurs for lower napin concentrations by using low methylated pectins like DM43. It confirms that interactions are mainly electrostatically driven.

No effect of DM has been observed neither for the foaming properties (foam formation and foam stability), nor for the thickness of foam films determined by SANS. The foam capacity, the foam stability and the foam film thickness are identical for both pectins used in this study.

The thickness of films in foams produced from pure napin solutions was 320 to 350 Å and around 315 Å in foams produced from napin/pectin mixtures, whatever the DM of pectins. SANS experiments indicated thereby that the thickness of films in the foams is more or less the same whatever the sample, pure napin or napin complexed to pectins. This let us think that free proteins is the only specie present within the films and contribute to the film foam stability. It has to be noticed that this value of thickness has already been found for CTAB/pectin systems by SANS [22] but also for synthetic polymers/surfactant systems by film disjoining pressure experiments [32, 33]. This may be explained by the fact that SANS measurements give thickness when the foams are in dry state, while disjoining pressure experiments give thickness under a high pressure applied on a single film. Both conditions could turn out to lead to the same thickness.

**Conclusions**

The impact of protein concentration, ionic strength and polymer charge density on the foaming properties has been analysed and can be related to the structures initially present in the bulk solutions (free proteins or complexes with different sizes) and finally in the foams.

For 1g/L of protein and 1g/L of pectin, at 25 mM of ionic strength, solutions are turbid and the size of complexes in the bulk reaches 1500 nm; 4% and 9% of free napin are present with respectively DM43 and DM74. The drainage was very fast in these conditions. At 148 mM of ionic strength, solutions are clear, the size of complexes equals 200 to 500 nm and 78% of free napin are present.

These results may be related to a recent study dealing with the impact of β-lactoglobulin aggregates on foaming properties [34]. Rullier et al. [34] observed that foaming properties are improved for systems containing protein aggregates compared to non aggregated solutions. The size of aggregates is also an important parameter: for small aggregates (35-71nm), foaming properties are better compared to free protein solutions, while for large aggregates (117-197nm), foaming properties are lower compared to free protein solutions. This study also highlighted that a mixture of both free proteins and aggregated proteins is required to obtain the most stable foams. The authors proposed the following mechanism: free proteins adsorb at the film surfaces and act as anchors for protein aggregates and lead to stable foam films if the amount of free proteins is sufficient. This mechanism of film stabilization is related to the formation of gel-like films as previously discussed for other systems [35, 36].

These conclusions could be applied with a parallel between protein aggregates in the study of Rullier et al. [34] and the napin/pectin complexes in our study.

In fact, we observed that foaming properties are improved for systems containing napin/pectin complexes compared to pure napin solutions. Smaller aggregates (200-500nm at 148mM) lead to form more stable foam; larger aggregates (1500nm at 25mM) lead to a fast drainage. A mixture of both free napins and napin/pectin complexes is required to obtain stable foams if the amount of free napin is sufficient (4% are not sufficient). It could be noted at this stage that the interfacial and rheological properties are also better in the mixtures.

These results enable us to clarify the role of each specie in the mixture, as crudely schematized in the graphic TOC:

- free proteins contribute to the foaming capacity, provided the initial free protein content in the bulk is sufficient to allow the foam formation

- soluble complexes slow down the drainage by their presence in the Plateau borders, which finally results in the stabilisation of foams.

**Acknowledgments**

This work has been supported in part by the Institut National de la Recherche Agronomique (INRA) and by the Region Pays de la Loire. Copenhagen Pectin is acknowledged for providing the Pectins. The Laboratoire Léon Brillouin, CEA, Saclay, is gratefully acknowledged for beam time allocation.